%% file: main.tex
\documentclass[10pt,twocolumn,letterpaper]{article}

\usepackage[pagenumbers]{cvpr} 

\input{preamble}

\definecolor{cvprblue}{rgb}{0.21,0.49,0.74}
\usepackage[pagebackref,breaklinks,colorlinks,allcolors=cvprblue]{hyperref}

\def\paperID{8188} 
\def\confName{CVPR}
\def\confYear{2025}

\title{OneNet: A Channel-Wise 1D Convolutional U-Net}

\author{
    Sanghyun Byun\(^{1,2\footnotemark[2]}\) \hspace{8mm}
    Kayvan Shah\(^1\) \hspace{8mm}
    Ayushi Gang\(^1\) \hspace{8mm}
    Christopher Apton\(^1\) \\
    Jacob Song \(^2\) \hspace{8mm}
    Woo Seong Chung \(^2\) \\
    \(^1\) University of Southern California \hspace{8mm}
    \(^2\) LG Electronics \\
    {\normalsize \tt \{byuns\footnotemark[2] , kpshah, agang, apton\}@usc.edu} \\
    {\normalsize \tt \{sang.byun, jaigak.song, wooseong.chung\}@lge.com}
}

\begin{document}
\maketitle
\footnotetext[2]{Corresponding author.}

\input{sec/0_abstract}

\input{sec/1_intro}

\input{sec/2_related}

\input{figures/3_method_overview}

\input{sec/3_method}

\input{figures/4_results_ablation}

\input{sec/5_results}

\input{sec/6_conclusion}

\hspace{2mm}

{\setlength{\parindent}{0cm}
\textbf{Acknowledgment} We thank the University of Southern California for providing AI computing resources and Dr. Yan Liu for professional support.
}

{
    \small
    \bibliographystyle{ieeenat_fullname}
    \bibliography{main}
}


\end{document}

%% file: preamble.tex
\usepackage[symbol]{footmisc}
\usepackage{algorithm}
\usepackage[noend]{algpseudocode}
\usepackage{multirow}
\usepackage{stfloats}
\fnbelowfloat


%% file: sec/0_abstract.tex
\begin{abstract}
Many state-of-the-art computer vision architectures leverage U-Net for its adaptability and efficient feature extraction. However, the multi-resolution convolutional design often leads to significant computational demands, limiting deployment on edge devices. We present a streamlined alternative: a 1D convolutional encoder that retains U-Net’s accuracy while enhancing its suitability for edge applications. Our novel encoder architecture achieves semantic segmentation through channel-wise 1D convolutions combined with pixel-unshuffle operations. By incorporating PixelShuffle, known for improving accuracy in super-resolution tasks while reducing computational load, OneNet captures spatial relationships without requiring 2D convolutions, reducing parameters by up to 47\%. Additionally, we explore a fully 1D encoder-decoder that achieves a 71\% reduction in size, albeit with some accuracy loss. We benchmark our approach against U-Net variants across diverse mask-generation tasks, demonstrating that it preserves accuracy effectively. Although focused on image segmentation, this architecture is adaptable to other convolutional applications. Code for the project is available at 
\href{https://github.com/shbyun080/OneNet}{https://github.com/shbyun080/OneNet}.

\end{abstract}

%% file: sec/1_intro.tex
\section{Introduction}
\label{sec:intro}

With advancements in encoder-decoder architectures, the accuracy and versatility of vision models have reached unprecedented levels. However, deploying these high-parameter models on edge devices (e.g., mobile phones) poses a significant challenge due to their limited computational resources. Techniques like optimization and quantization have become essential to enable the use of state-of-the-art models on these devices. 

Many modern architectures, including diffusion models \cite{freeu, harmonyview, multidepth}, rely heavily on the U-Net \cite{unet} architecture as an encoder-decoder backbone. Yet, its structure is not optimized for efficiency in resource-constrained environments as they often overlook the inherent architectural inefficiencies. Since they typically employ a standard convolutional backbone such as ResNet \cite{resnet}, the parameter count can escalate rapidly, impacting efficiency and increasing the chance of overfitting. Although the architecture is highly adaptable, minimal research has been conducted to streamline its size for edge deployment. To solve this issue, we propose modifying the U-Net \cite{unet} backbone to reduce the number of parameters, thereby decreasing computing costs and model download size. This optimization would make U-Net \cite{unet} more feasible for edge deployment and open up possibilities for more complex models by reallocating resources to tasks of greater importance.

\input{figures/3_method_block}

In contrast, areas like image super-resolution have long benefited from techniques like PixelShuffle \cite{Pixelshuffle}, significantly improving pipeline efficiency without compromising spatial information. However, despite the clear advantages of these scaling techniques, they have not been widely explored in other domains. Additionally, while lightweight architectures like MobileNet \cite{mobilenet} have been effective for more straightforward tasks such as classification, such structures remain underutilized in generative models. This knowledge gap suggests an opportunity to explore alternative efficiency-driven architectures for demanding vision tasks, potentially unlocking new performance levels and adaptability in model deployment.

This paper proposes a novel adaptation of the U-Net \cite{unet} architecture that bridges the gap between state-of-the-art performance and edge-deployability by reducing the model size while preserving accuracy. Our approach is the first to leverage channel-wise 1D convolutions in conjunction with pixel-shuffling operations to enable efficient feature and spatial attention without reliance on 2D computations. By eliminating these operations that are often challenging to parallelize on resource-constrained edge devices, we ease the burden on sequential computing cores and make the model more suitable for lightweight deployment. Additionally, we optimize spatial processing by reducing kernel sizes, focusing instead on cross-feature interactions to further minimize memory requirements. This novel architecture can seamlessly replace the standard U-Net in existing pipelines, offering a versatile and high-efficiency solution for edge applications. Our major contributions are as follows:
\begin{itemize}
    \item We design a novel convolution block that only uses 1D convolutions while retaining spatial information through the introduction of pixel-unshuffle downscaling, pixel-shuffle upscaling, and channel-wise 1D convolutions.
    
    \item We implement two versions of U-Net \cite{unet} (1D encoder with 2D-decoder and 1D encoder-decoder) using our novel convolution block, effectively reducing the model size by 47\% and 67\% while maintaining reasonable accuracy. 

    \item We evaluate our proposed method variations on multiple mask-prediction datasets and compare our results to commonly used backbones for U-Net to display performance retention while reducing the model's total size and the number of computations.
\end{itemize}

%% file: figures/3_method_block.tex
\begin{figure}[!t]
\centering
     \begin{subfigure}[b]{\columnwidth}
        \centering
        \includegraphics[width=.99\textwidth]{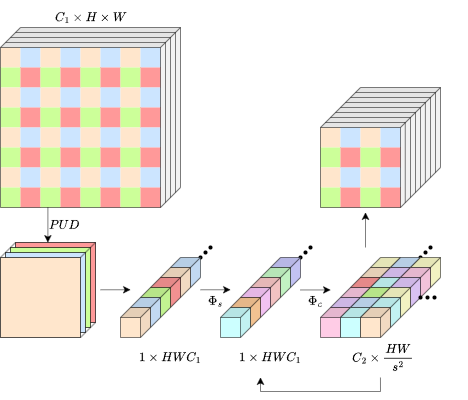}
        \caption{}
        \label{fig:ps-a}
    \end{subfigure}%
    
    \begin{subfigure}[b]{\columnwidth}
        \centering
        \includegraphics[width=.99\textwidth]{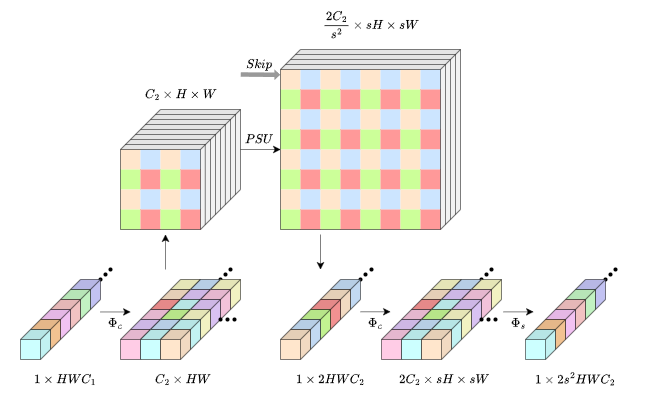}
        \caption{}
        \label{fig:ps-b}
    \end{subfigure}%
    \caption{\textbf{Channel-Wise 1D Convolution Block} (a) Encoder convolution block with pixel-unshuffle downscaling replacing max pooling operation, followed by a single spatial and two channel-wise layers. (b) Decoder convolution block with pixel-shuffle upscaling for tensor upsampling, followed by a spatial layer between two channel-wise layers.}
\label{fig:block}
\end{figure}

%% file: sec/2_related.tex
\section{Related Works}
\label{sec:related}

\textbf{U-Net} \cite{unet, unet++} introduces a new encoder-decoder approach to semantic segmentation, employing a technique of a contractive path followed by an expansive path. Developed for smaller mask-count datasets, it displays significant improvement over the existing state-of-the-art networks \cite{deepNN, resnet}. One advantage of U-Net \cite{unet} is its ability to use a variety of backbones \cite{resnet, mobilenet, skipinit, eunnet} depending on the complexity of the task. U-Net \cite{unet} and its variations \cite{unet++, attentionunet} have become a cornerstone in the medical imaging field, especially for MRI image segmentation. Gupta et al. \cite{MRI-SA} utilizes U-Net \cite{unet} for brain tumor segmentation with notable success. Similarly, SAM-2 \cite{sam2} extends this approach by introducing spatiotemporal mask predictions for videos. In contrast, Zhuang et al. \cite{tdc} propose a novel idea to tackle the challenge of the lack of segmentation labels in video data by using a temporally dependent classifier (TDC) to mimic the human-like recognition procedure. However, despite these advances, standard U-Net \cite{unet} architectures remain computationally demanding for edge devices.

\textbf{1D Convolutional Neural Networks} \cite{1dmri, 1dtime, conv1d} have been widely used for various special applications such as spectral MRI \cite{1dmri} and time-series \cite{1dtime} problems. As 1D calculations require fewer computations in most cases, they are preferred in most dimension-separable tasks. Additionally, recent research by Kirchmeyer et al. \cite{conv1d} demonstrates that a ConvNet consisting entirely of directional 1D convolutions performs comparably on ImageNet classification. Such results indicated that 1D convolutions can achieve similar performances on further complex tasks such as segmentation.

\textbf{Pixel Shuffle} \cite{Pixelshuffle} increases image resolution by converting low-resolution (LR) tensors into high-resolution (HR) images. The operation takes an $W \times H \times Cr{^2}$ tensor and converts it into a high-resolution $Wr \times Hr \times C$ tensor. In the field of image super-resolution, multiple state-of-the-art methods \cite{imagesr-ntire, shuffleunet} propose leveraging PixelShuffle \cite{Pixelshuffle} for efficient processing of images at a lower resolution without losing higher-resolution information. Our work is largely inspired by this technique, as it helps capture spatial relationships while reducing computational load.

\textbf{Efficient segmentation models} \cite{mobileseg, blitzmask, mobilevig, mobilenet, lidar} have seen a rise in demand, leading to the development of models such as PP-MobileSeg \cite{mobileseg}, Blitzmask \cite{blitzmask}, or Mobilevig \cite{mobilevig}, which are designed specifically for mobile devices. Although PP-MobileSeg \cite{mobileseg} achieves real-time segmentation on lightweight architectures, it struggles when dealing with large datasets that require high spatial attention. Our approach addresses this limitation by employing a state-of-the-art technique that simplifies spatial attention without sacrificing efficiency. Similarly, TFNet \cite{lidar} focuses on fast and accurate segmentation for LiDAR data, which aligns with our overall aim of developing efficient, high-performance models for resource-constrained environments.





%% file: figures/3_method_overview.tex
\begin{figure*}[ht]
\centering
    \includegraphics[width=\textwidth]{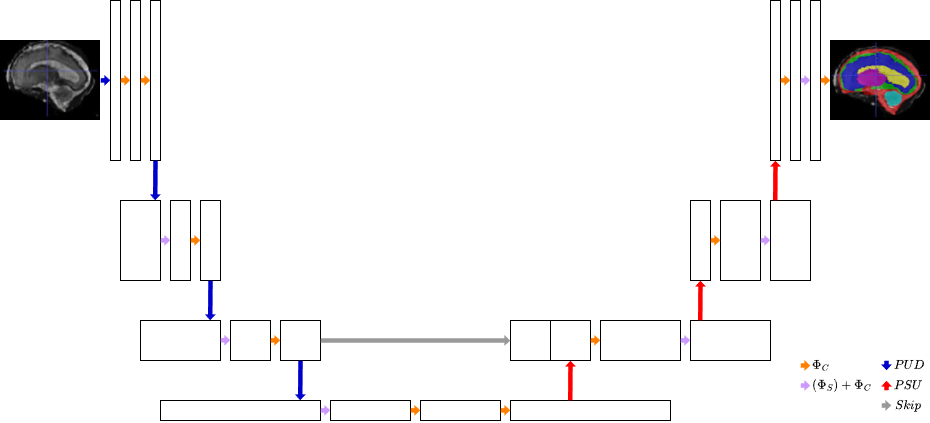}
    \caption{\textbf{Channel-Wise 1D Encoder-Decoder} OneNet employs a U-Net \cite{unet} architecture with skip connections for segmentation tasks. The architecture above is a 3-layer variant shown for simplicity. The encoder block replaces the max pool layer with pixel-unshuffle downscaling, with the image downscaled immediately on input for spatial relations to be captured. The decoder block replaces upsampling methods with a pixel-shuffle upscaling. In the architecture shown, 1D convolution is used for both encoder and decoder, with optional spatial convolution. To satisfy the spatial-preservation property in the decoder, we only decode to half resolution. The top layer of the decoder is implemented without batch normalization or ReLU to avoid zero-centering of the prediction head.}
\label{fig:overview}
\end{figure*}

%% file: sec/3_method.tex
\section{OneNet}
\label{sec:method}

We stated that our approach heavily relies on pixel-shuffling operations for the preservation of spatial information. This motivates us to treat the channel dimension as a spatial-augmented feature vector. We first cover the basis of pixel-unshuffle downscaling and its parameter-reduction property in \cref{sec:method:shuffle}. Then, we explain our implementation of 1D convolutional blocks used to leverage the unique downsampling technique in \cref{sec:method:conv}. Lastly, we elaborate on how U-Net \cite{unet} can be adapted to the proposed OneNet architecture through a breakdown of the encoder (\cref{sec:method:encoder}) and the decoder (\cref{sec:method:decoder}).

\subsection{Pixel-Unshuffle Downscaling}
\label{sec:method:shuffle}
We express the downsampling operation as a function $Y(m)=X(D(mS))$, where $m$ is the output tensor index of $Y$, $m$ is the total number of pixels in $Y$, $S$ stands for the scaling factor, and $D$ is the sampling strategy used, such as max pooling. The target of downsampling is to increase the receptive field of the network while decreasing the computation cost. Traditional convolutional neural networks achieve this through either max pooling or average pooling.

We propose to replace the pooling operation with pixel-unshuffle downscaling, which can be annotated as 
\begin{equation}
\label{eq:pud}
D(X)_{i,a,b}=X_{\lfloor i/s^2 \rfloor, sa+\lfloor sa/i \rfloor,sb+sb(mod \hspace{1mm} i)}
\end{equation}
where the $D$ denotes pixel-unshuffle downscaling operation, $X$ is the input tensor, $s$ is the scale of the downscaling, and $i \in [0,s^2C)$ where $C$ is the number of channels in $X$. Although a scale greater than 2 is feasible, the resolution of the tensor is reduced by the factor of $s^L$ where $L$ is the number of layers. Thus, it would not leave sufficient room for downscaling for most datasets as $s=3$ and $L=4$ would downscale by the factor of $81$ compared to $s=2$ of $16$.

For ease of discussion, we simplify the tensor with a single spatial dimension and a single channel dimension. We start with a tensor, with spatial information on column and channel on row dimension, as shown below
\begin{equation}
X=
\begin{pmatrix}
a_{00} & a_{01} \\
a_{10} & a_{11}
\end{pmatrix}
\end{equation}

With the traditional 2D approach where the convolutional layer has a kernel size of $k=(2,2)$ and stride of $s=1$ to produce $X'$, and the max pool layer $D$ is applied with the assumption that the first element has the largest value, we get $X(D(mS))=(a'_{00}, a'_{10})^T$. The number of multiplication calculations here would be the product of parameter size ($kC=4$), number of convolutions ($MS/s=2$), and number of output features $C=2$, resulting in $kMSC^2=16$ operations.

In comparison, the proposed method would first perform pixel-unshuffle downscaling to attain $X'=(a'_{00},a'_{01},a'_{10},a'_{11})^T$. Then, a 1D convolution is applied to directly reduce the number of channels to attain
\begin{equation}
X(D(mS))=
\begin{pmatrix}
W_0X' \\
W_1X'
\end{pmatrix}
\end{equation}
which would result in the kernel size of $MC=4$ and the number of output features $C=2$, resulting in a total multiplication count of $MC^2=8$. This is half of the multiplication counts needed compared to the 2D approach.

\input{figures/3_method_conv}
\label{sec:method:conv}

\subsection{1D-Kernel Convolution}
\label{sec:method:kernel}
Ignoring batch dimension for the sake of simplicity, traditional convolution block performs a 3D computation in spatial and channel dimensions, resulting in a parameter of $(C_{in}, H, W, C_{out})$, where $C$ is the channel, $H$ is the height, and $W$ is the width of the input tensor. In depth-wise convolution proposed by Howard et al. \cite{mobilenet}, this weight is decomposed into spatial and channel dimensions, resulting in the parameter of $(H,W)+(C_{in}, C_{out})$.

Although MobileNet \cite{mobilenet} implementation significantly reduces the number of parameters, it fails to capture spatial-channel dependencies due to its design. We show the convolutional steps taken by each method in \cref{fig:conv}.

As pixel-unshuffle downscaling transfers spatial knowledge to the channel axis, we replace the 2D convolution layer with 1D convolution layers working on a flattened tensor of size $(B, HWC)$. Channel-wise convolution processes this tensor and runs $C_{out}$ 1D convolutions with the kernel size and stride of $k=s=C_{in}$ to attain a tensor of size $(B, C_{out}, HW)$, effectively achieving parameter size of $(C_{in}, C_{out})$. As spatial information is still intact in the channel dimension, running a channel-wise convolution results in consideration of spatial-channel dependencies.

As spatial information solely depends on pixel-unshuffle downscaling, the scale factor also defines the model's receptive field, which is $ Receptiveness=SL$. In the U-Net \cite{unet} architecture with OneNet encoder, the model would depend on the decoder for a larger receptive field.

\input{figures/3_method_pud_algo}

\input{figures/4_results_overall}

\input{figures/4_results_param}

\subsection{1D Encoder}
\label{sec:method:encoder}

We show our proposed OneNet encoder block in \cref{fig:ps-a}, consisting of pixel-unshuffle downscaling, followed by two channel-wise 1D convolutions with an optional spatial convolution. One issue with pixel-unshuffling implementation is the required chain of flattening and shape-restoration steps. Thus, we propose a method for preserving all tensors in a single spatial dimension by implementing a 1D-compatible pixel-unshuffle algorithm. We show our adaptation in \cref{algo:pud}.

To compare the parameter size of the traditional 2D convolution block to the proposed OneNet encoder block, let us start with a tensor $X$ of shape $(C, H, W)$ after a downsampling operation for the 2D encoder. With the 2D encoder, following \cref{fig:conv-1d}, the two convolutions would have weight of shapes $(C, k, k, 2C)$ and $(2C, k, k, 2C)$. With a stride of 1 and a kernel size $k=3$, this results in multiplication counts of $2k^2C^2HW$ and $4k^2C^2HW$, totaling $6k^2C^2HW$ calculations. For OneNet, the input tensor would have the shape of $(4C,H,W)$ due to pixel-unshuffle downscaling, giving weights of shapes $(4C, 2C)$ and $(2C, 2C)$. This gives the total multiplication count of $8C^2HW+4C^2HW=12C^2HW$. Thus, the ratio of parameters for a single block would be $(6k^2C^2)/(12C^2)=k^2/2$. As $k$ must be at least 2, this would result in a smaller computation and parameter size.

Additionally, we test an optional spatial convolution that works on a flattened 1D tensor of size $(CHW)$ with a kernel size $k<<C$ and stride $s=1$. We do not test our model on a pre-trained encoder as per U-Net \cite{unet} implementation. As the paper focuses on the backbone potential of 1D convolutions, pre-training is out of the scope of our paper.

\subsection{Segmentation Decoder}
\label{sec:method:decoder}
Additionally, we design and propose a decoder block (\cref{fig:ps-b}) that performs upsampling blocks through pixel-shuffle upscaling. Unlike the OneNet encoder, we perform an upscaling operation, which moves information from channel to spatial instead. Thus, the spatial information is not much represented in the channel dimension. Despite this difference, if the OneNet decoder is used in conjunction with the proposed encoder, we can make a mild assumption that the channel dimension already consists of sufficient spatial information due to the unshuffling steps performed. However, this signifies that although the encoder can be used in conjunction with any decoder architecture, the OneNet decoder is strictly specific to the OneNet encoder.

Otherwise, the decoder follows a standard U-Net \cite{unet} architecture with skip connections with a similar block as the proposed decoder. As the upscaling method reduces the channel size to one-fourth, there is no need to change the channel count after the concatenations from skip connections.

%% file: figures/3_method_conv.tex
\begin{figure}[!t]
\centering
     \begin{subfigure}[b]{\columnwidth}
        \centering
        \includegraphics[height=2cm]{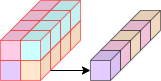}
        \caption{}
        \label{fig:conv-2d}
    \end{subfigure}%

    \hspace{1mm}
    
    \begin{subfigure}[b]{\columnwidth}
        \centering
        \includegraphics[height=2cm]{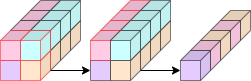}
        \caption{}
        \label{fig:conv-mobile}
    \end{subfigure}%
    
    \hspace{1mm}
    
    \begin{subfigure}[b]{\columnwidth}
        \centering
        \includegraphics[height=2cm]{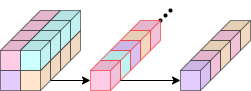}
        \caption{}
        \label{fig:conv-1d}
    \end{subfigure}%
    \caption{\textbf{Comparison of Convolutional Block} (a) Traditional 2D convolutional block with max pooling. (b) MobileNet \cite{mobilenet} block with max pooling. (c) OneNet implementation with pixel-unshuffle downscaling followed by 1D convolution.}
\label{fig:conv}
\end{figure}

%% file: figures/3_method_pud_algo.tex
\begin{algorithm}[t]
\caption{1D pixel-unshuffle downscaling}\label{alg:cap}

\textbf{Input: } $X$ in shape $(B,C,H \times W)$, height H, width W
\textbf{Output: } pixel-unshuffled $X$ in shape $(B,H \times W \times C)$
\begin{algorithmic}

\State $I \gets []$
\For{$i \gets 0$ to $\frac{W}{2}$}
    \For{$j \gets 0$ to $\frac{H}{2}$}
        \State $t \gets 2i+2j$
        \State $I \gets [t, t+1, t+w, t+w+1]$
    \EndFor
\EndFor

\State $X \gets X[:, :, I].reshape(B,C,-1,4)$
\State $X \gets X.transpose(0,2,1,3).flatten(dim=1)$

\end{algorithmic}
\label{algo:pud}
\end{algorithm}

%% file: figures/4_results_overall.tex
\setlength\tabcolsep{2.4pt}
\renewcommand{\arraystretch}{1.8}
\begin{table*}[t]
\centering
\scriptsize
\begin{tabular}{|l*{6}{|*{3}{c}}|}
\hline
    \multirow{2}{*}{\textbf{Method}} & \multicolumn{3}{|c|}{VOC \cite{voc}} & \multicolumn{3}{|c|}{PET$_F$ \cite{pet}} & \multicolumn{3}{|c|}{PET$_S$ \cite{pet}} & \multicolumn{3}{|c|}{MSD Heart \cite{msd}} & \multicolumn{3}{|c|}{MSD Brain \cite{msd}} & \multicolumn{3}{|c|}{MSD Lung \cite{msd}}\\
    & $\mathcal{L}_{CE}$ & mAP$_{0.5}$ & mIOU & $\mathcal{L}_{CE}$ & mAP$_{0.5}$ & mIOU & $\mathcal{L}_{CE}$ & mAP$_{0.5}$ & mIOU & $\mathcal{L}_{CE}$ & mAP$_{0.5}$ & mIOU & $\mathcal{L}_{CE}$ & mAP$_{0.5}$ & mIOU & $\mathcal{L}_{CE}$ & mAP$_{0.5}$ & mIOU\\  
\hline
    U-Net$_{4}$ \cite{unet}     & 1.985 & 0.0541 & 0.182 & 2.206 & 0.0329 & 0.316 & 0.243 & 0.7717 & 0.713 & 0.0223 & 0.4305 & 0.063 & 0.0455 & 0.2450 & 0.001 & 0.0010 & 0.5542 & \textbf{0.009} \\
    ResNet$_{34}$ \cite{resnet} & 1.321 & 0.0806 & 0.332 & \textbf{0.648} & 0.0503 & 0.597 & 0.179 & 0.9098 & 0.801 & 0.0087 & 0.4375 & 0.065 & 0.1305 & 0.4188 & 0.079 & 0.0010 & 0.5529 & 0.009 \\
    ResNet$_{50}$ \cite{resnet} & \textbf{1.079} & \textbf{0.0921} & \textbf{0.372} & 1.027 & \textbf{0.0512} & \textbf{0.599} & \textbf{0.189} & \textbf{0.9303} & \textbf{0.815} & 0.0086 & 0.4368 & 0.065 & 0.0496 & 0.5419 & 0.010 & \textbf{0.0007} & 0.5566 & 0.009 \\
    MobileNet \cite{mobilenet}  & 2.007 & 0.0480 & 0.166 & 2.386 & 0.0329 & 0.252 & 0.262 & 0.6091 & 0.664 & 0.0288 & 0.3265 & 0.047 & \textbf{0.0351} & 0.5764 & 0.011 & 0.0010 & \textbf{0.5787} & 0.008 \\
\hline
    OneNet$_{e, 4}$             & 2.144 & 0.0485 & 0.160 & 2.713 & 0.0279 & 0.216 & 0.309 & 0.5176 & 0.636 & \textbf{0.0041} & \textbf{0.4396} & \textbf{0.066} & 0.0363 & \textbf{0.5789} & \textbf{0.105} & 0.0007 & 0.5531 & 0.009 \\
    OneNet$_{ed, 4}$            & 2.553 & 0.0363 & 0.149 & 3.080 & 0.0227 & 0.172 & 0.437 & 0.2179 & 0.535 & 0.0076 & 0.4187 & 0.062 & 0.0455 & 0.5415 & 0.099 & 0.0009 & 0.5510 & 0.008 \\
\hline
\end{tabular}
\caption{\textbf{Baseline Comparisons on Semantic Segmentation Datasets} U-Net and OneNet are trained on datasets \cite{voc, pet, msd} without pre-training for a fair comparison, as outlined by the original U-Net \cite{unet}. ResNet \cite{resnet} and MobileNet \cite{mobilenet} are pre-trained on ImageNet-1K for accuracy comparison. PET$_F$ and PET$_S$ stand for the full and small mask versions of the Oxford Pet \cite{pet} dataset with 38 and 3 classes each. U-Net$_i$ stands for vanilla U-Net \cite{unet} encoder with downsampling layers of $i$. Resnet$_i$ stands for $i$-layer version of ResNet \cite{resnet}. OneNet$_{e(d), i}$ has $i$ downsampling layers, with $e$ and $d$ each standing for encoder and decoder replaced with the proposed architecture, respectively. The best results are in \textbf{bold}.}
\label{results:overall}
\end{table*}

%% file: figures/4_results_param.tex
\setlength\tabcolsep{5pt}
\renewcommand{\arraystretch}{1.6}
\begin{table*}
\centering
\scriptsize
\begin{tabular}{|l|cccc|}
\hline
    \textbf{Method} & \# Param (M) & Param (MB) & FLOPS (GB) & Memory (MB) \\
\hline
    U-Net$_{4}$ \cite{unet} & 31.04 & 124.03 & 104.72 & 509.61 \\
    U-Net$_{5}$ \cite{unet} & 124.42 & 497.41 & 130.80 & 524.29 \\
    ResNet$_{34}$ \cite{resnet} & 25.05 & 98.07 & 29.40 & \textbf{241.17} \\
    ResNet$_{50}$ \cite{resnet} & 74.07 & 287.83 & 84.98 & 450.36 \\
    MobileNet \cite{mobilenet} & 14.40 & 57.47 & 83.96 & 671.09 \\
\hline
    OneNet$_{e, 4}$ & 16.39 & 65.42 & 78.42 & 639.63 \\
    OneNet$_{e, 5}$ & 65.73 & 262.63 & 98.82 & 656.41 \\
    OneNet$_{ed, 4}$ & \textbf{9.08} & \textbf{36.30} & \textbf{22.92} & 799.01 \\
    OneNet$_{ed, 5}$ & 36.38 & 145.47 & 39.00 & 885.00 \\
\hline
\end{tabular}
\caption{\textbf{Comparison on Model Size} Number of parameters (in millions), parameter size, FLOPS used, and memory used during inference is reported. With the 4-layer model, OneNet achieves 47\% reduction with encoder swap and 69\% reduction with encoder-decoder swap. A sample tensor of size $(1, 3, 256, 256)$ was used as the network input. The best results are shown in \textbf{bold}. U-Net$_i$ stands for vanilla U-Net \cite{unet} encoder with downscaling layers of $i$. Resnet$_i$ stands for $i$-layer version of ResNet \cite{resnet}. OneNet$_{e(d), i}$ has $i$ downscaling layers, with $e$ and $d$ each standing for encoder and decoder, respectively.}
\label{results:param}
\end{table*}

%% file: figures/4_results_ablation.tex
\setlength\tabcolsep{5pt}
\renewcommand{\arraystretch}{1.8}
\begin{table*}[t]
\centering
\scriptsize
\begin{tabular}{|l*{3}{|*{4}{c}}|}
\hline
    \multirow{2}{*}{\textbf{Method}} & \multicolumn{4}{|c|}{PET$_S$ \cite{pet}} & \multicolumn{4}{|c|}{MSD Heart \cite{msd}} & \multicolumn{4}{|c|}{MSD Lung \cite{msd}} \\
    & $\mathcal{L}_{CE}$ & mAP$_{0.5}$ & mIOU & DSC & $\mathcal{L}_{CE}$ & mAP$_{0.5}$ & mIOU & DSC & $\mathcal{L}_{CE}$ & mAP$_{0.5}$ & mIOU & DSC \\  
\hline
    OneNet$_{e, 4}$-S          & 0.320 & \textbf{0.5275} & 0.632 & 0.963 & \textbf{0.0034} & 0.4383 & 0.065 & \textbf{0.699} & \textbf{0.0006} & 0.5510 & \textbf{0.008} & \textbf{0.115} \\
    OneNet$_{e, 4}$-C          & \textbf{0.309} & 0.5176 & \textbf{0.636} & \textbf{0.967} & 0.0363 & \textbf{0.4386} & \textbf{0.066} & 0.605 & 0.0007 & \textbf{0.5531} & 0.009 & \textbf{0.115} \\
    OneNet$_{ed, 4}$-S         & 0.416 & 0.2253 & 0.529 & 0.868 & 0.0125 & 0.3938 & 0.059 & 0.544 & 0.0007 & 0.5512 & \textbf{0.008} & 0.112 \\
    OneNet$_{ed, 4}$-C         & 0.437 & 0.2179 & 0.535 & 0.842 & 0.0076 & 0.4187 & 0.062 & 0.605 & 0.0007 & 0.5510 & \textbf{0.008} & 0.111 \\
\hline
\end{tabular}
\caption{\textbf{Ablation Study of Spatial Convolution} We study the effect of spatial convolution before channel convolution for datasets with small segmentation mask counts \cite{pet, msd}. The best results are in \textbf{bold}. We see an overall performance to be unaffected by the addition of spatial convolution with a kernel size of 9 for both encoder-only and encoder-decoder replacements.}
\label{results:ablation}
\end{table*}

%% file: sec/5_results.tex
\section{Experiments}
\label{sec:results}

We implement the OneNet architecture using PyTorch \cite{pytorch} and optimize it with the Adam \cite{adam} optimizer, setting an initial learning rate of $1 \times 10^{-4}$. We set a learning rate scheduler decaying the learning by a factor of $0.1$ every 20 epochs starting from epoch 50. The model is trained for a total of 300 epochs, with batch sizes of 32 or 64 depending on memory availability, using an input image resolution of $512 \times 512$ and a segmentation mask resolution of $256 \times 256$. We omit batch normalization and ReLU activation from the top decoder layer to ensure accurate segmentation predictions, avoiding any potential zero-centering effects on the output. The initial bottleneck channel count is set to 64, while the kernel size for the optional spatial convolution layer is set to 9. The pixel-unshuffling downscaling is applied with a scale factor of 2, resulting in $16\times$ total downscaling factor for a 4-layer encoder. Training is conducted on a single RTX 4090 GPU with 24GB of VRAM.

\subsection{Mask Prediction Tasks}

We evaluate our method across six diverse datasets: PASCAL VOC \cite{voc}, Oxford Pet with breed masks and subject masks \cite{pet}, and three medical segmentation datasets—MSD Heart, MSD Brain, and MSD Lung \cite{msd}. These datasets represent a broad spectrum of masking complexities, with class counts ranging from 38 in Oxford Pet with breed masks to as few as 2 for MSD Lung, and intermediate counts of 21, 3, 4, and 4 for the other datasets. Consistent with Ronneberger et al. \cite{unet}, we do not pre-train the encoder for both U-Net \cite{unet} and OneNet models to provide a controlled comparison, allowing us to assess OneNet's performance without the advantages of transfer learning. However, for models such as  ResNet \cite{resnet} and MobileNet \cite{mobilenet}, which typically benefit from large-scale pre-training, we initialize them on ImageNet-1K \cite{imagenet} and then fine-tune them on each specific dataset. This provides a nuanced view of OneNet’s capabilities in comparison to commonly used models in the field. Our accuracy evaluation, detailed in \cref{results:overall}, uses U-Net \cite{unet} as a primary baseline to ensure a fair and relevant comparison across architectures.

\textbf{Training setup and evaluation metric} As all datasets explored are multi-class segmentation tasks, cross-entropy loss is used with the weight of the background mask reduced to a quarter. We report cross-entropy loss ($\mathcal{L}_{CE}$), mean average precision with an IOU threshold of 0.5 (mAP$_{0.5}$), and mean intersection-over-union (mIOU).

\textbf{Analysis on medical tumor segmentation}
For all medical tumor detection datasets \cite{msd}, \cref{results:overall} indicates that the proposed OneNet model performs on par with established baseline architectures. Notably, for the MSD Heart dataset \cite{msd}, OneNet shows a slight edge in accuracy, achieving a 2\% improvement over the baseline models. This suggests that OneNet can maintain competitive performance even in specialized applications, capturing important features required for medical imaging.

Additionally, OneNet achieves model size reduction while maintaining an accuracy drop within 1\% across datasets, accompanied by a substantial 47\% reduction in model parameters. This underscores OneNet’s ability to perform complex segmentation tasks with a far more compact architecture. Such efficiency makes it particularly well-suited for applications involving smaller-mask prediction tasks, including medical segmentation and tasks requiring precise depth or normal map generation.

The ability to sustain high accuracy with fewer parameters means that OneNet is optimized for both performance and resource constraints, making it an excellent candidate for real-time or edge-based medical imaging applications. By minimizing computational demands without a significant compromise in accuracy, OneNet offers a practical, scalable solution for high-precision tasks in medical diagnostics and other specialized segmentation applications.

\textbf{Analysis on general multi-class segmentation} 
The results in \cref{results:overall} show an 11\% and 15\% drop in accuracy for the PASCAL VOC \cite{voc} and full-size Oxford Pet \cite{pet} datasets, respectively, with a 10\% decrease for the Oxford Pet dataset with fewer masks for encoder-only replacement. These declines highlight a trade-off between accuracy and model size, especially when working with many classes. However, we consider this trade-off acceptable to enable edge deployability where resources are limited.

This pattern suggests that the receptive field created in OneNet encoder-decoder by pixel-unshuffle downscaling (with a scale factor of 2) effectively enhances local feature detection rather than image-wide classification. Thus, encoder-only replacement would be better suited for such tasks requiring a large receptive field, supporting the two proposed models' efficiency without overly compromising performance, making it suitable for resource-constrained deployments.

\subsection{Impact on Model Size}
In \cref{results:param}, we present the parameter size and FLOPs of OneNet compared to baseline models \cite{unet, resnet, mobilenet}, using a sample input tensor of size $(1, 3, 256, 256)$ for baselines and $(1, 3, 512, 512)$ for OneNet. OneNet achieves substantial efficiency gains, with approximately 47\%/25\% reductions in parameters and FLOPs for its encoder and 67\%/78\% reductions for the complete encoder-decoder structure. These outcomes align closely with the theoretical calculations in \cref{sec:method:shuffle} and demonstrate OneNet's capability to handle high-resolution inputs while significantly reducing computational demands.

The compact size of OneNet is particularly noteworthy, as it can reach below 40MB—a 71\% reduction compared to the standard U-Net. This drastic reduction makes OneNet highly suitable for deployment on edge devices, allowing the model weights to be downloaded and stored locally with minimal storage overhead. However, we observe an increase in memory usage due to the pixel-unshuffle downscaling, which increases the channel count by a factor of 4 compared to the constant channel dimension in max pooling. While this higher channel count imposes some memory cost, it enhances the model's ability to capture fine-grained spatial information, ultimately improving feature representation without compromising overall efficiency.

\subsection{Ablation on Spatial Convolutions}
Further study of the effect of channel convolutions in the OneNet convolution blocks is shown in \cref{results:ablation}. To discuss whether spatial convolution similar to that of a MobileNet \cite{mobilenet} would influence the model, we add the spatial convolutions to the 4-layer encoder and encoder-decoder OneNet replacements as shown in \cref{fig:overview}, with kernel size 9 and stride 1. We additionally report the Dice coefficient (DSC) for pixel-wise agreement. The results suggest that due to the pixel-unshuffle downscaling's organization of the pixels in a convolutional manner without repetition, such grouping drives the spatial layer to have minimal effect. This supports channel-wise convolution's sufficiency in capturing spatial information as the addition of a spatial module, albeit small, does not affect the model's accuracy in any way.

Further analysis of the role of channel-wise convolutions within OneNet's convolution blocks is presented in \cref{results:ablation}. To examine whether spatial convolutions, similar to those in MobileNet \cite{mobilenet}, would impact model performance, we incorporate spatial convolutions into the OneNet 4-layer encoder and encoder-decoder structures, as depicted in \cref{fig:overview}, using a kernel size of 9 and stride of 1. Alongside this configuration, we report the Dice coefficient (DSC) to measure pixel-wise agreement and assess segmentation accuracy.

With an accuracy difference in the range of 2\%, the results indicate that the pixel-unshuffle downscaling technique—which organizes pixels in a structured convolutional manner without redundancy—minimizes the need for spatial convolutions. This design allows the channel-wise convolution to effectively capture spatial information independently. The addition of the spatial convolution module had a negligible impact on accuracy, underscoring that OneNet’s channel convolutions are sufficient for spatial feature extraction. Thus, these findings validate the efficiency of OneNet’s architecture by demonstrating that a lightweight, channel-focused approach is capable of high spatial accuracy without the added complexity of a spatial convolution layer.

%% file: sec/6_conclusion.tex
\section{Conclusion}
\label{sec:conclusion}

We presented OneNet, a novel architecture that effectively reduces the number of parameters required while preserving accuracy through the use of 1D convolutions. Our approach excels in local-feature-centric mask-generating tasks by relying solely on channel-wise 1D convolutions, ensuring that spatial attention is maintained through the use of pixel-unshuffle downscaling. In contrast to existing models that require large parameter counts due to an increasing number of channels and constant kernel sizes, our method leverages channel-wise features to significantly reduce model size without compromising performance. We demonstrate that pixel-unshuffle downscaling is both efficient and information-preserving, allowing for stable feature transfer between scales without the need for traditional 2D convolutions. Additionally, we show that the concept of pixel shuffling can be effectively applied to decoder networks by assuming spatial dependencies. Looking forward, we are keen to explore various pixel-repositioning techniques across different architectures, aiming to further advance the development of efficient AI models that can deliver high performance with reduced computational overhead.